\documentclass{aa}  
\usepackage{graphicx}
\usepackage{txfonts}
\usepackage{subcaption}
\usepackage{hyperref}

\newcommand{\feh}{{[Fe/H]}\xspace}
\newcommand{\gaia}{{\it Gaia}\xspace}
\newcommand{\kms}{{$\rm{km}\,\rm{s}^{-1}$}\xspace}
\newcommand{\ks}{{$K{\rm s}$}\xspace}

\begin{document} 
\defcitealias{Verberne_2024}{V24}

   \title{Double progenitor origin of the S-star cluster}

   \author{Sill Verberne\inst{1}\fnmsep\thanks{Corresponding author: Sill Verberne\\
          \email{verberne@strw.leidenuniv.nl}}
          \and
          Elena Maria Rossi\inst{1}
          \and 
          Sergey E.~Koposov\inst{2,3}
          \and
          Zephyr Penoyre\inst{1}
          \and 
          Manuel Cavieres\inst{1}
          \and 
          Konrad Kuijken\inst{1}
          }

      \institute{Leiden Observatory, Leiden University,
              P.O. Box 9513, 2300 RA Leiden, the Netherlands
         \and
             Institute for Astronomy, University of Edinburgh, Royal Observatory, Blackford Hill, Edinburgh EH9 3HJ, UK
         \and
             Institute of Astronomy, University of Cambridge, Madingley Road, Cambridge CB3 0HA, UK
             }

   \date{Received 24 February 2025 / Accepted 31 March 2025}

  \abstract
  {The origin of the cluster of S-stars located in the Galactic Centre is tied to the supermassive black hole Sagittarius A*, but exactly how is still debated. In this paper, we investigate whether the Hills mechanism can simultaneously reproduce both the S-star cluster's properties and the observed number of hypervelocity stars. To do so, we forward-modelled the capture and disruption of binary stars originating from the nuclear star cluster (NSC) and the clockwise disc (CWD). We find that the ratio of evolved to main-sequence S-stars is highly sensitive to the origin of the binaries, and that neither the injection of binaries from the CWD nor from the NSC exclusively can reproduce all observations. However, when considering the injection of binaries from both locations, we are able to reproduce all the observations simultaneously, including the number of observed hypervelocity stars, the evolutionary stage of the S-stars, their luminosity function, and the distribution of their semi-major axes. The implications  of our findings  include that $\sim90\%$ of hypervelocity stars ejected over the past $\sim10$ Myr should originate from the CWD, that the main-sequence S-stars originated in the CWD, and that the evolved S-stars originated in an old stellar population such as the NSC.}

   \maketitle

\section{Introduction}
The Galactic Centre (GC) is unique for its combination of relative proximity and extreme stellar dynamics. It is, however, also a challenging environment to study due to high line-of-sight extinction and source crowding. The GC is host to the super-massive black hole Sagittarius A* (Sgr A*) with a mass of $4.3\times10^6$ M$_\odot$ \citep[][]{Eisenhauer_2005, Ghez_2008, Schodel_2009, Gillessen_2009, Boehle_2016, Gillessen_2017, Do_2019, Gravity_2019, GRAVITY_2024} and a nuclear star cluster (NSC). In addition, Sgr A* is orbited by a cluster of mainly B-type main-sequence stars out to $\sim0.04$ pc called the S-star cluster \citep[see review by][]{Genzel_2010}. The origin of this cluster remains unknown, since the strong tidal force from Sgr A* inhibits standard star formation from molecular clouds in this region \citep{Genzel_2010}. Nonetheless, the S-star cluster has been intensely studied, due to the possibility of tracing individual orbits of stars and thereby constraining the mass and distance to Sgr A* \citep[e.g.][]{Ghez_2008, Gillessen_2009, DO_2013, Gillessen_2017, Gravity_2018}.

Within the highly complex environment of the GC, exotic gravitational interactions can occur. Of particular interest to this work is the Hills mechanism, in which a stellar binary approaches a massive black hole (MBH) to within the tidal radius where the tidal force of the MBH overcomes the self gravity of the binary \citep{Hills_1988, Yu_2003}. The result is that one star is captured into a tight elliptical orbit, while the other is ejected as a hypervelocity star (HVS). HVSs can travel at $\gtrsim10^3$ \kms, making them unbound to the Galactic potential \citep{Hills_1988, Bromley_2006, Rossi_2014}. The captured stars on the other hand have been suggested as a possible formation mechanism for the S-star cluster, since they should get deposited at radii similar to those of the S-stars \citep[e.g.][]{Gould_2003, Ginsburg_2006}. In addition, recent observations have provided strong indirect evidence that the Hills mechanism operates in the GC \citep{Koposov_2020}. 

In this study, we examine whether the observed properties of both the S-star cluster and HVSs are consistent with predictions by the Hills mechanism. This will tell us whether the entire S-star cluster might have formed through the Hills mechanism or whether a more complex assembly history is required to explain the observed properties of the S-stars. What makes this study unique compared to earlier studies \citep[e.g.][]{Madigan_2009, Zhang_2013, Generozov_2020} is that we simultaneously simulated both the S-star and HVS populations and compared both with state-of-the-art observations.

In Section~\ref{sec:S-stars}, we summarise our observational knowledge of the S-star cluster. Section~\ref{sec:method} describes our methods for forward modelling the population of S-stars and HVSs. In Section~\ref{sec:progenitor_populations}, we investigate two possible origins for progenitor binaries. The discussion on our results and potential future work are presented in Section \ref{sec:discussion}. Finally, in Section~\ref{sec:conclusions} we give our conclusions.

\section{The S-star cluster}\label{sec:S-stars}
We start by giving an overview of the observed properties of the S-star cluster in Section~\ref{sec:overview}, before discussing the effects of observational selection in Section~\ref{sec:selection_function}.

\subsection{Overview}\label{sec:overview}
The S-star cluster has isotropically distributed stellar orbits with a radial extent of roughly $\sim0.04$ pc from Sgr A*; it is formed by mainly B-type massive main-sequence stars, and it is surrounded by at least one disc of massive young stars containing many O/WR stars out to $\sim0.5$ pc \citep{Levin_2003, Paumard_2006, Bartko_2009, Fellenberg_2022}. No O/WR stars are detected in the inner 0.04 pc \citep{Genzel_2010}.

The S-star cluster has been observed for nearly three decades \citep{Eckart_1996, Ghez_1998}. During this time, a range of instruments have been used to study its properties with techniques including adaptive optics-assisted astrometry, integral field spectroscopy, and optical/infrared interferometry. Due to the high line-of-sight extinction \citep[$A_{\rm V}\sim50$;][]{Fritz_2011}, the study of the S-star cluster is done around 2 $\mu$m, where extinction is lower \citep[$A_{\rm Ks}\sim2.5$;][]{Schodel_2010, Fritz_2011}. 

The number of parameters needed to fully describe an orbit around the GC is six. Usually, these are sky position, proper motion, radial velocity, and acceleration in the plane of the sky \citep{Gillessen_2017}. A catalogue containing orbits for 40 stars is published in \citet{Gillessen_2017}, most of which are in the central arcsecond. Furthermore, a catalogue of 36 young stars with orbital solutions is given in \citet{Fellenberg_2022}. More recently, orbital parameters were determined for 20 stars at 2--7 arcsec from Sgr A* in \citet{Young_2023}.

Observations indicate that S-stars follow a thermal eccentricity distribution ($n(e)\propto e$) and travel on orbits with periods as short as decades, with the shortest observed period being about $13$ yr \citep{Gillessen_2017}. The most studied S-star is S2: a bright ($m_{\rm K}=14$, without extinction correction), massive star ($M_{\rm ZAMS}\sim14$ M$_\odot$) on a 16 year orbit \citep[e.g.][]{Habibi_2017}. The metallicity of the S-stars is currently unknown \citep{Habibi_2017}.

The young disc surrounding the S-star cluster, often referred to as the clockwise disc (CWD), has a top-heavy initial mass function (IMF) with a slope of around $-1.7$ \citep{Lu_2013, Gallego-Cano_2024}. The CWD is consistent with a single star formation episode about 2--6 Myr ago \citep{Paumard_2006}. There is also evidence for additional discs at varying radii \citep[e.g.][]{Fellenberg_2022}. 

\begin{figure}
    \centering
    \includegraphics[width=\linewidth]{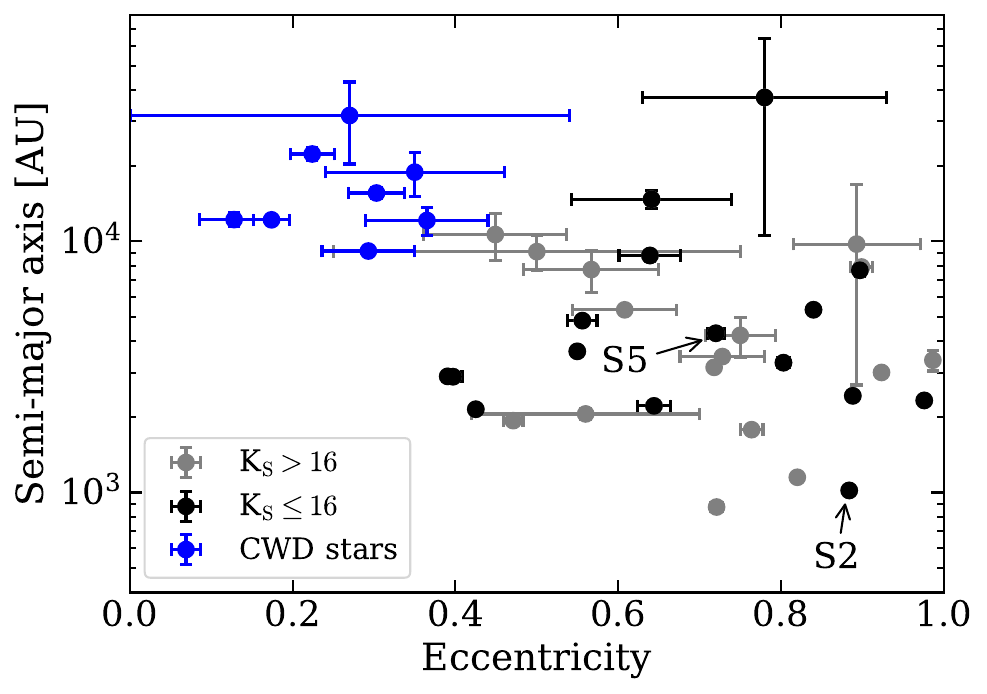}
    \caption{Orbital solutions for stars at the GC found in \citet{Gillessen_2017}, with the exception of S5, which is from \citet{Fellenberg_2022}. Stars that are considered part of the young disc are coloured blue, which are S66, S67, S83, S87, S91, S96, S97, and R44. S111 is omitted, because of the hyperbolic solution. The grey points show S-stars fainter than our completeness limit, while black points show those that are brighter.}
    \label{fig:S-stars}
\end{figure}

In Fig.~\ref{fig:S-stars}, we show orbital solutions in the 2D space of eccentricity against semi-major axis, for stars belonging to either the CWD or the S-star cluster \citep[data from][]{Gillessen_2017}. Our figure clearly shows that these two stellar structures occupy different ranges of this parameter space.

\subsection{Observational selection function}\label{sec:selection_function}
The stars for which orbital parameters have been determined cannot be seen as an unbiased dataset for the properties of the S-star cluster: accelerations are more difficult to measure for faint stars with relatively long orbital periods. To understand this bias, we need a prescription of the observational selection function. Using mock observations, \citet{Burkert_2024} finds an empirical relation that connects a star's pericentre distance and eccentricity to the chance that an orbital solution would have been found in \citet{Gillessen_2017}. However, \citet{Burkert_2024} did not consider the apparent magnitude of a star to be part of the selection function. Since they base their selection on our ability to measure the acceleration in the plane of the sky within their mock observations, it is implicitly assumed that a radial velocity measurement was also available. To account for this additional factor, we assumed the orbital solutions to be complete down to \ks$=16$, based on the detection limit of {\it SINFONI} for a typical run \citep{Gillessen_2009}. Since the \citet{Gillessen_2017} catalogue, only a single additional orbital solution for a \ks$<16$ star within the \citet{Burkert_2024} completeness limit has been published \citep[S5;][]{Fellenberg_2022}. This star was reported as being consistent with the CWD, but given its high eccentricity and small semi-major axis (typical for S-stars) we instead treated it as an S-star. For \ks$<16$, there are respectively twelve and four orbital solutions for main-sequence and evolved S-stars in our dataset. We did not consider S111, since it has a hyperbolic orbital solution \citep{Gillessen_2017}. Throughout this work, our stellar magnitudes are not corrected for extinction.

\section{Method}
\label{sec:method}
Now that we have discussed the properties and observations of the S-star cluster, we will describe how the Hills mechanism could explain the formation of both the S-star cluster and HVSs. We briefly review the stellar dynamics underlying this model in Section~\ref{sec:hills}, and discuss the implementation of this theory in our simulations in Section~\ref{sec:simulations}. 

\begin{figure}
    \centering
    \includegraphics[width=\linewidth]{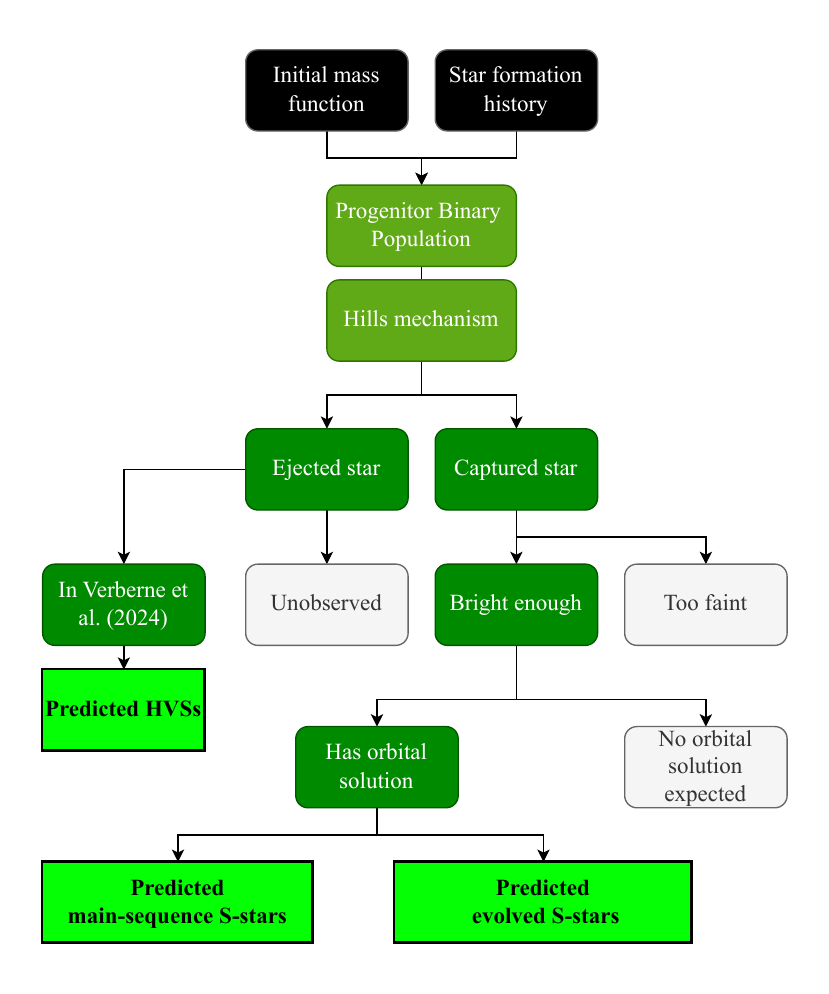}
    \caption{Schematic overview of the steps of our forward model that predicts the numbers of observed HVSs, main-sequence S-stars, and evolved S-stars.}
    \label{fig:diagram}
\end{figure}

A summary of our approach is as follows. For a given progenitor binary star formation history, IMF, and Hills mechanism rate, we predicted the number of observed main-sequence and evolved S-stars, the semi-major axes of their orbits around Sgr A*, and their luminosity function. Additionally, we predicted the number and properties of the related HVSs observable in the \citet[][hereafter V24]{Verberne_2024} survey. Fig.~\ref{fig:diagram} shows a schematic overview of our modelling procedure. Finally, these predictions are simultaneously compared to i) the observed properties of the S-star cluster, and ii) the number of observed HVSs in the \citetalias{Verberne_2024} survey.

\subsection{Hills mechanism}\label{sec:hills}
In the Hills mechanism \citep{Hills_1988}, a stellar binary approaches a massive black hole within the tidal radius, where the tidal force of the black hole separates the binary, ejecting one star as an HVS and capturing the other star in a tight, elliptical orbit \citep[for a review, see][]{Brown_2015}. The captured stars have been suggested as the possible origin of the S-stars \citep[e.g.][]{Gould_2003, Ginsburg_2006}. 

\begin{figure*}
    \centering
    \includegraphics[width=\linewidth]{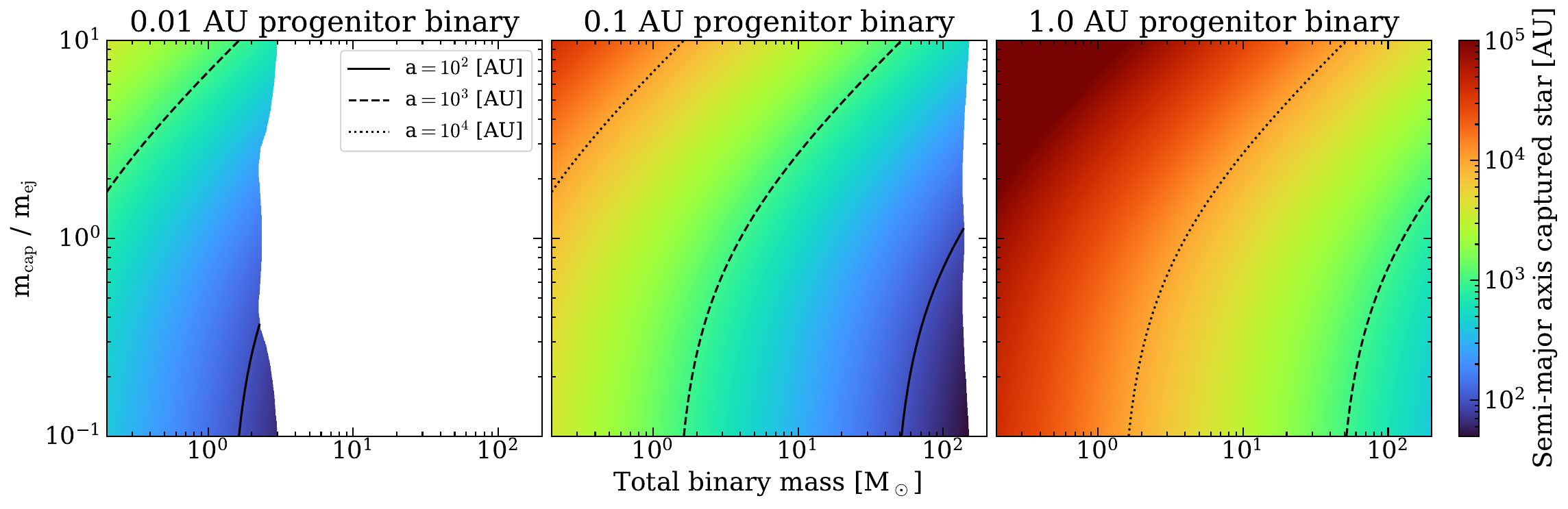}
    \caption{The semi-major axis of the captured star's orbit  around Sgr A* as a function of the progenitor binary's total mass and mass ratio, for three different binary separations; the mass ratio is defined as the mass of the captured star over that of ejected star. We only show solutions for which the progenitor binary separation is larger than the sum of the stellar radii at zero-age main-sequence.}
    \label{fig:a_vs_q}
\end{figure*}

Progenitor binaries are expected to be on approximately parabolic orbits \citep{Kobayashi_2012}, for which the ejection probability is equal for both stars and does not depend on the mass ratio \citep{Sari_2010}. In this three-body encounter, there is an energy exchange, where this amount 
\begin{equation}\label{eq:delta_E}
\Delta E = \alpha^2\frac{m_{\rm ej}m_{\rm cap}G}{a_{\rm b}}\left(\frac{M_{\rm bh}}{m_{\rm b}}\right)^{1/3} \quad {\rm and} \quad m_{\rm b} = m_{\rm cap}+m_{\rm ej},
\end{equation}
is gained by the ejected star at the expense of the captured one.
In the expression for $\Delta E$, $\alpha$ is a prefactor of order unity that depends on the geometry of the encounter (the inclination of the binary orbital plane, the binary phase at closest approach, the pericentre radius, and the binary eccentricity), $m_{\rm ej}$ and $m_{\rm cap}$ are the mass of the ejected and captured star respectively, $G$ the gravitational constant, and $a_{\rm b}$ the semi-major axis of the progenitor binary. Thus the ejection velocity of the ejected star is
\begin{equation}\label{eq:Vej}
V_{\rm ej} = \sqrt{\frac{2\Delta E}{m_{\rm ej}}}=\alpha\sqrt{\frac{2Gm_{\rm cap}}{a_{\rm b}}}\left(\frac{M_{\rm bh}}{m_{\rm b}}\right)^{1/6}
\end{equation}
\citep[e.g.][]{Rossi_2014}. The captured star, on the other hand, will travel on an orbit with a semi-major axis equal to 
\begin{equation}
a_{\rm cap} = \frac{G M_{\rm bh} m_{\rm cap}}{2\Delta E} =\frac{a_{\rm b}M_{\rm bh}}{2\alpha^2m_{\rm ej}}\left(\frac{m_{\rm b}}{M_{\rm bh}}\right)^{1/3}
\end{equation}
after the separation of the binary system. In Fig.~\ref{fig:a_vs_q}, we show the semi-major axis of the captured star as a function of the total binary mass and the mass ratio of the captured over the ejected star, where we take $\alpha=1$.
This figure clearly demonstrates that Hills mechanism disruptions of binaries will result in stars captured into orbits with semi-major axes similar to those of the observed S-stars.

The initial pericentre distance ($r_{\rm p}$) of the captured star is equal to the pericentre distance of the binary around the MBH and is set by the penetration factor ($\beta$) and tidal radius ($r_{\rm t}$) defined as
\begin{equation}\label{eq:beta}
    \beta=\frac{r_{\rm t}}{r_{\rm p}} \quad {\rm and} \quad  r_{\rm t} = a_{\rm b}\left(\frac{M_{\rm bh}}{m_{\rm b}}\right)^{1/3}.
\end{equation} 
We can thus write the initial eccentricity of the captured star as 
\begin{equation}
e_{\rm init} = 1- \frac{r_{\rm t}}{\beta a_{\rm cap}} = 1 - \frac{a_{\rm b}}{\beta a_{\rm cap}}\left(\frac{M_{\rm bh}}{m_{\rm b}}\right)^{1/3}.
\end{equation}
We refer to this pericentre distance and eccentricity as initial, because the angular momentum of a star deposited in the S-star cluster is expected to change rapidly due to dynamical resonant relaxation processes \citep[e.g.][]{Rauch_1996, Hopman_2006, Binney_2008, Antonini_2013, Generozov_2020}. The timescale on which the energy of the orbit changes is much longer and happens on the two-body relaxation timescale. The precise timescales on which the angular momentum and energy change are still an active field of research \citep[see review by][]{Alexander_2017}. Additionally, stellar collisions might play an important role near Sgr A* \citep[e.g.][]{Sari_2019, Rose_2023, Rose_2024, Ashkenazy_2024}.

\subsection{Simulations} \label{sec:simulations}
Having discussed the theory behind the Hills mechanism and its potential to explain the S-star cluster, we now describe our simulations.

In our model, progenitor binaries are characterised by mass, age, metallicity, mass ratio, and binary separation. We assumed stars with solar metallicity \citep[Z$_{\odot}=0.0142$;][]{Asplund_2009} for simplicity and evaluated the impact of this assumption in Section~\ref{sec:metallicity}. Other properties of our progenitor binaries were drawn from probability distributions. In particular, the distribution of the mass ratio is $f(q)\propto q^\gamma$, where $0.1\leq q\leq1$ and we used an orbital period ($P$) distribution of the form $\log(P/[1 \rm\ sec])\propto (\log P/P_0)^\pi$, where $P_0$ is 1 sec. Our prior ranges are $-2\leq\gamma\leq2$ and $0 \leq\pi\leq2$. The mass and age of the binaries were determined given a star formation history and an initial mass function (IMF). Using the star formation history and the IMF, we can compute the 'present-day' mass function at any look-back time. For a Hills mechanism disruption at look-back time $t$, we calculated the number of stars alive per unit mass $m$ and unit age $A$ from
\begin{equation}\label{eq:mAt}
    \frac{dN}{dm dA}(m,\ A\ |\ t) = {\rm SFR}(A+t)\ C(A)\ P(m|A),
\end{equation}
where ${\rm SFR}(A+t)$ is the star formation rate at time $A+t$, $C(A)$ the fraction of stars alive at age $A$ given an IMF, and $P(m\ |\ A)$ the mass function for stars alive of age $A$, which is the IMF truncated at the maximum mass of a star alive with age $A$. If a star survives till the present day, it has an age $A+t$. The fraction of stars alive at age $A$ is calculated by
\begin{equation}\label{eq:CA}
    C(A) = \int P(m) \left(m < m_{\rm max}(A)\right) dm,
\end{equation}
where $P(m)$ is the normalised IMF, and $m_{\rm max}(A)$ the maximum mass star alive at age $A$. We assumed the rate of binary disruptions at a point in time is proportional to the number of stars alive at that time. The Hills mechanism rate we report is always the current rate.

\begin{figure}
    \centering
    \includegraphics[width=\linewidth]{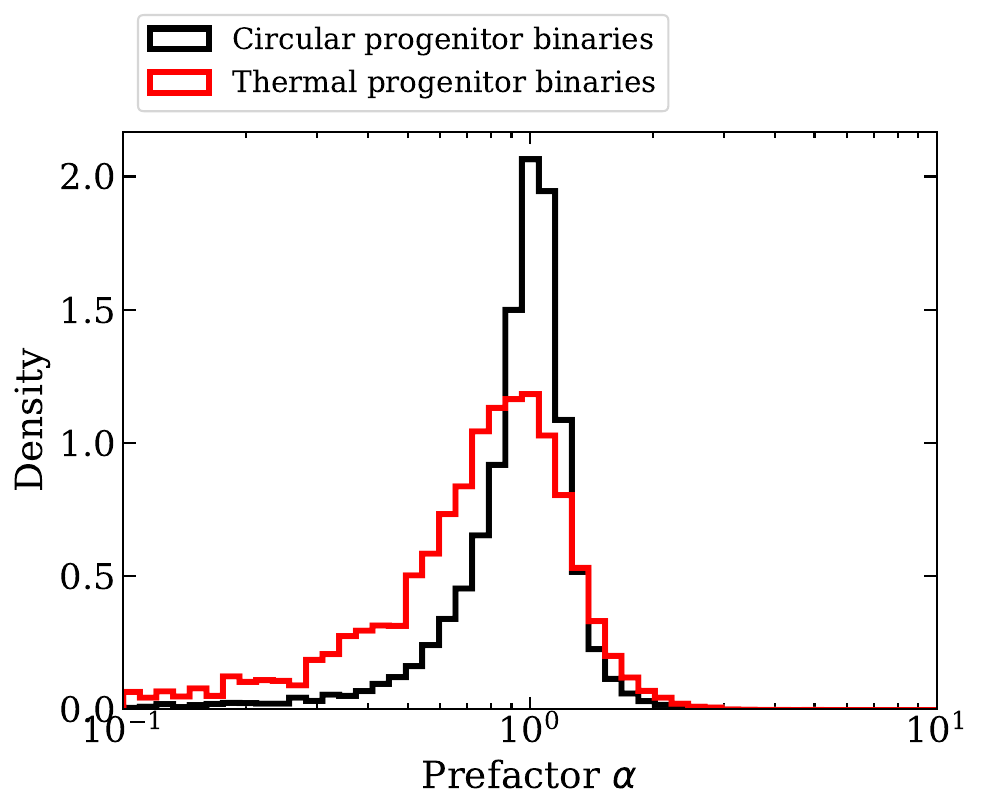}
    \caption{Prefactor $\alpha$ that we sample over to account for random orbital phases and inclinations at the time of disruption (Sersante et al.~in prep.) The black and red lines show the distribution for circular and thermally distributed progenitor eccentricities respectively.}
    \label{fig:alpha}
\end{figure}

Unless stated otherwise, we assumed that the progenitor binaries are disrupted isotropically, which means that the resulting ejected and captured stars have random orientations. We calculated the energy that is exchanged in the tidal separation from equation~\ref{eq:delta_E}. We used the results from the simulations in Sersante et al.~(in prep.) and randomly sampled over the prefactor $\alpha$, using the distribution shown in Fig.~\ref{fig:alpha}.
We assumed here that the progenitor binaries are on circular orbits in the full lose-cone regime \citep{Merritt_2013} such that $\beta$ (equation~\ref{eq:beta}) is distributed log-uniformly. We also investigated a thermal eccentricity distribution in Section~\ref{sec:eccentricity}. 

\subsubsection{S-star predictions}
\label{sec:s-star_predictions}
After the binary is disrupted, we calculated if the captured star is within the stellar tidal disruption radius of Sgr A* \citep{Rees_1988}. If so, the star was removed from the simulation, if not we assumed that the eccentricity is rapidly thermalised such that $n(e)\propto e$ while the orbital energy remains constant. The reason for this is that simulations point out that resonant relaxation can thermalise the S-star eccentricity distribution over $\sim10^7$ yr \citep[e.g.][]{Perets_2009, Antonini_2013, Generozov_2020}, which is of the order of the age of the CWD. The extended mass required for resonant relaxation to be fast enough is similar to the $1\sigma$ upper limit on the enclosed mass within the S2 star orbit \citep{Generozov_2020, GRAVITY_2024}. After the orbits were thermalised, we again removed any stars whose pericentre lies within the stellar tidal disruption radius of Sgr A*. On the other hand, we kept the semi-major axes of stars fixed, because we assumed negligible changes in the orbital energy over the timescales of interest (see Section~\ref{sec:E_diff}). We followed the stellar evolution of the captured stars using \texttt{MIST} until the present epoch and determined the \ks magnitude \citep{Mist0, MistI}. We assumed a uniform foreground extinction of $A_{K \rm s}=2.5$ \citep{Schodel_2010, Fritz_2011} towards the GC and corrected the photometry accordingly. 

We only investigated stars with \ks$<16$, because of observational completeness (see Section~\ref{sec:selection_function}). We assigned to each remaining star a probability of having a known orbital solution following \citet{Burkert_2024}. We used the equivalent evolutionary point (EEP) number in \texttt{MIST} to separate main-sequence stars from evolved stars at the terminal-age main-sequence (EEP $=454$). This leaves us with a number of predicted main-sequence and evolved S-stars that we can compare to observations in \citet{Gillessen_2017}.

\subsubsection{HVS predictions}\label{sec:hvs_predictions}
Simultaneously to the analysis of the captured stars presented above, we tracked the ejected stars using \texttt{Speedystar}\footnote{\url{https://github.com/fraserevans/speedystar}} \citep{Rossi_2017, Marchetti_2018, Contigiani_2019, Evans_2022_I, Evans_2022_II}. This code simulates the ejection, propagation, and evolution of HVSs and provides synthetic photometry and \gaia observables for simulated HVSs. Our procedure is identical to that presented in \citetalias[][section~4.3]{Verberne_2024}, except that we now considered realistic star formation histories as described in Section~\ref{sec:progenitor_populations}. This was incorporated through the prescription in equations~\ref{eq:mAt} and \ref{eq:CA}. 

\citetalias{Verberne_2024} performed a dedicated survey of HVSs. They rely on the near-radial trajectories of HVSs to identify candidates and provide follow-up observations. The survey has a clearly defined selection function, allowing us to assess the detectability of any given simulated star. \citetalias{Verberne_2024} identify a single previously known HVS, S5-HVS1 \citep{Koposov_2020}: a 1700 \kms star at about 9 kpc from us that originates in the GC, where it was ejected about 4.8 Myr ago. This is the only star unambiguously linked to a GC ejection and provides strong indirect evidence of the Hills mechanism.

For a given star formation history, we forward modelled the expected number of HVSs in the dedicated survey presented in \citetalias{Verberne_2024}. The forward modelling leaves us with a predicted (simulated) population of HVSs that we can compare to the one HVS found observationally in \citetalias{Verberne_2024}. Therefore, when we mention the predicted number of HVSs, we are referring to the predicted number of HVSs in the \citetalias{Verberne_2024} survey. The main constraining power of these observations is in providing an upper limit to the Hills mechanism rate, since the detection of only one HVS is very informative. 

We consider HVSs to be the best comparison point since, following their definition in \citetalias{Verberne_2024}, they are almost certainly formed through the Hills mechanism. Although different formation mechanisms might contribute to the observed S-star population, we know that Hills mechanism disruptions will put stars into S-star-like orbits (Section~\ref{sec:hills}).

\section{Progenitor populations}\label{sec:progenitor_populations}
In this section we will investigate two specific progenitor populations for binaries disrupted through the Hills mechanism.

\subsection{First scenario: binaries originated from the NSC}
\label{sec:NSC}
We start by considering if binaries originating from parsec scales in the NSC (half light-radius around $\sim 5-10$ pc) can explain the observed properties of the S-star cluster and HVSs.

The origin of progenitor binaries on which the Hills mechanism operates is uncertain. If stellar relaxation dominates the influx of disrupted binaries, the binaries will tend to originate at or beyond the sphere of influence ($\sim 3$ pc) of the MBH \citep{Perets_2007}. Additionally, Penoyre et al.~(in prep.) find that the axi-symmetry of the central potential of the Milky Way produces a collisionless flux of binaries from around the sphere of influence radius that dominates over the collisional rate. Thus, it appears unavoidable that stellar binaries from the NSC are set on orbits that result in their tidal separation, although the rate is still uncertain. We therefore assumed here that all binaries originate from the NSC and have the same star formation history as the NSC. 

\begin{figure}
    \centering
    \includegraphics[width=\linewidth]{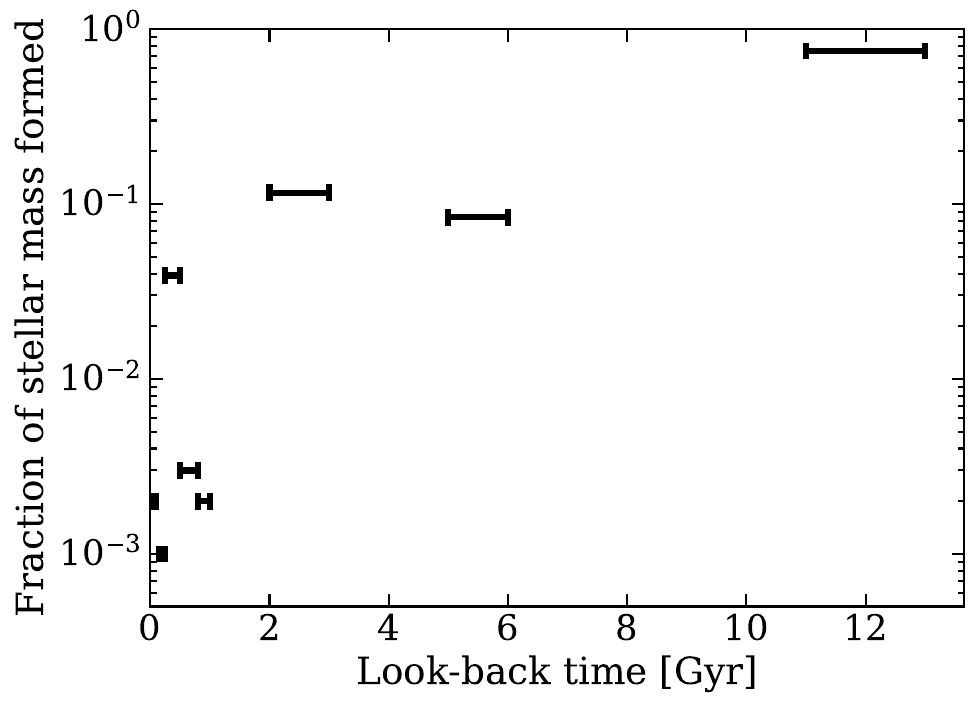}
    \caption{Look-back time against the fraction of stellar mass formed over each star formation epoch, which encapsulates the star formation history of the nuclear star cluster as determined in \citet{Schodel_2020} for a $2Z_\odot$ model. The bars indicate the bin sizes of \citet{Schodel_2020} within which we assume the star formation rate was constant.}
    \label{fig:SFH_NSC}
\end{figure}

The NSC stellar population is consistent with having a canonical IMF \citep{Kroupa_2001}. On the other hand, its star formation history is still debated, but it is known that it has not been constant \citep[e.g.][]{Pfuhl_2011, Schodel_2020, Chen_2023}, with most of the population having formed billions of years ago. The model of the star formation history we used is the super-solar metallicity ($2Z_{\odot}$) model excluding the very young stars in the CWD from \citet[][table 3]{Schodel_2020}. In Fig.~\ref{fig:SFH_NSC}, we plot the fraction of the total stellar mass formed as a function of look-back time.
We can see that the NSC is dominated by old stars, with only a few percent having been formed in the last billion years. We assumed that the star formation rate was constant within the intervals reported in \citet{Schodel_2020}.

We only considered captured stars with $M>0.8$ M$_{\odot}$, since lower mass stars are not bright enough to be observed near the GC at \ks$<16$. Evolved 1 M$_{\odot}$ stars, on the other hand, might still be observed at those magnitudes according to \texttt{MIST}. The sample is dominated by relatively low-mass binaries, since the bulk of the population is very old. 

\begin{figure*}
    \centering
    \begin{subfigure}{0.49\textwidth}
    \includegraphics[height=0.77\linewidth]{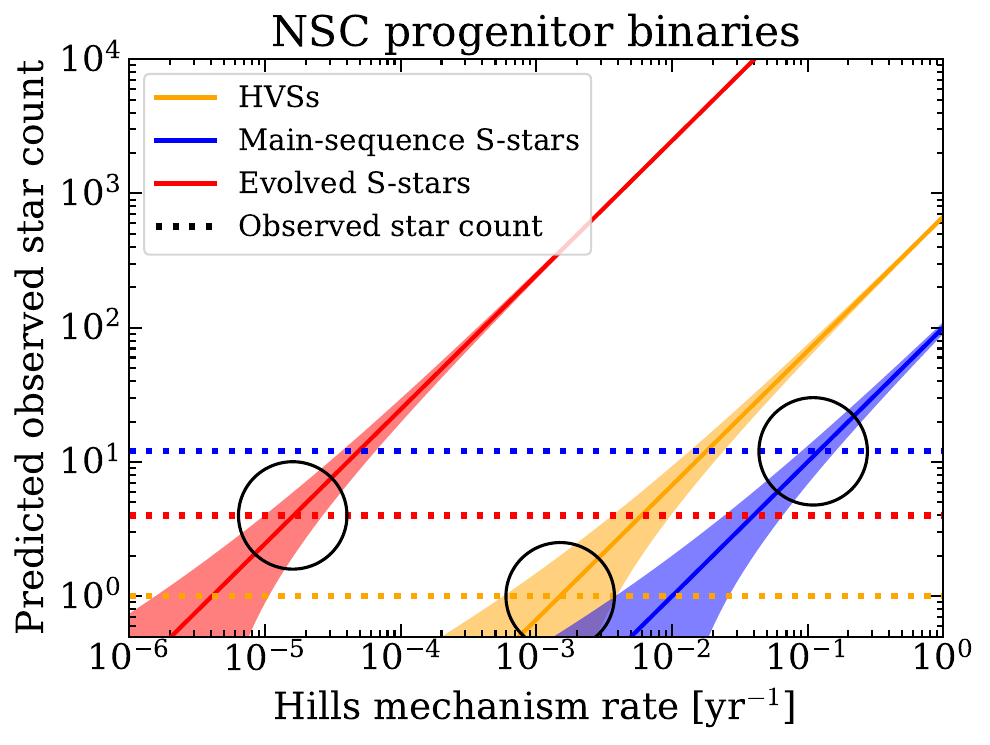}
    \end{subfigure}
    \begin{subfigure}{0.49\textwidth}
    \includegraphics[height=0.77\linewidth]{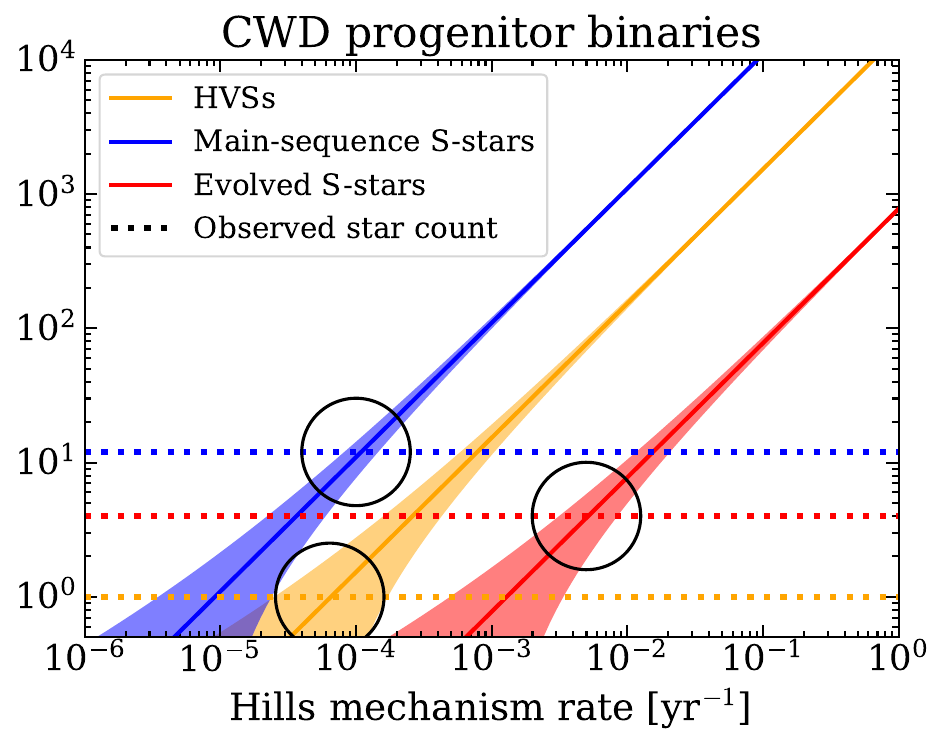}
    \end{subfigure}
    \caption{Hills mechanism rate against the predicted number of observed HVSs, main-sequence S-stars, and evolved S-stars. The dashed horizontal lines correspond to the observed number in each of these categories. The black circles highlight where the model and data agree. {\bf Left:} the model predictions when progenitor binaries for the Hills mechanism originate from the NSC. {\bf Right:} same as the left panel, but for binaries originating from the young CWD.}
    \label{fig:rate_num}
\end{figure*}

In the left panel of Fig.~\ref{fig:rate_num}, we show our results for binaries originating from the NSC.
Since the predicted numbers linearly scale with the Hills mechanism rate, their ratio remains constant: for every observed HVS we expect 0.15 and $3.7\times10^2$ main-sequence and evolved S-stars respectively within our brightness limit. This is incompatible with observations because 
\begin{itemize}
    \item for every observable main-sequence S-star, around $2.5\times10^3$ evolved S-stars should be visible unless they are somehow depleted (see Section~\ref{sec:collisions}),
    \item to account for the number of observed main-sequence S-stars, the Hills mechanism would deplete the NSC in $\sim100$ Myr \citep[$M_{\rm NSC}\sim2\times10^7 M_\odot$;][]{Feldmeier-Krause_2017},
    \item and there should be $\sim10^2$ HVSs in the \citetalias{Verberne_2024} survey footprint.
\end{itemize}

\subsection{Second scenario: binaries originated from the CWD}\label{sec:CWD}
We showed in the previous section that Hills mechanism disruptions of binaries originating from the overall population of the NSC cannot simultaneously explain the observed properties of the S-star cluster and HVSs. Here we will investigate an alternative progenitor binary population, originating from the CWD \citep[see also][]{Madigan_2009, Madigan_2014, Koposov_2020, Generozov_2020, Generozov_2021}. This alternative scenario is largely inspired by the overlap between the orbital plane of the progenitor binary of S5-HVS1 around Sgr A* and that of the CWD \citep[see Fig.~10 in][]{Koposov_2020}, which is suggestive since HVSs are ejected in the plane of the progenitor binary around the MBH. The age of the CWD is uncertain, but likely less than 10 Myr \citep{Lu_2013}.

We assumed a single star formation episode 10 Myr ago, and we modeled it with a Gaussian distribution centred at that epoch, with a standard deviation of 1 Myr, and additionally a top-heavy IMF with a slope of $-1.7$ \citep{Lu_2013, Gallego-Cano_2024}. Only stars more massive than about 7.5 M$_\odot$ are bright enough to pass our brightness limit given the age of the cluster. A clear difference with respect to the NSC scenario is the relative abundance of massive binaries, due to the younger ages of the stars and their top-heavy IMF.

Because HVSs are ejected in the plane of the progenitor binary around the MBH, we included this selection effect in our analysis of the expected number of HVSs in the \citetalias{Verberne_2024} survey. This was done by only considering HVS ejections in the plane of the CWD, described by an inclination and longitude of the ascending node of $(i, \Omega) = (130^\circ, 96^\circ)$, and a half-width at half-maximum of about $15^\circ$ \citep{Yelda_2014}. For a description of this coordinate system, see \citet{Lu_2009}.

In the case of the NSC, we assumed that the progenitor binaries were on parabolic orbits around the MBH. Because the CWD is much closer to Sgr A*, the centre of mass of the progenitor binary might approach SgrA* on a trajectory with significantly lower total (negative) energy (i.e. lower eccentricity). To account for this, we sampled from the radial surface density of the CWD \citep[$\sim r^{-2}$; e.g.][]{Fellenberg_2022} between 1 and 8 arcseconds from Sgr A*. The sampled semi-major axis sets the initial energy of the binary. We then calculated the semi-major axis of the captured star around Sgr A* taking into account this initial energy. In practice, this has a limited impact, since the initial energy is much smaller than the exchanged energy by a factor of $\sim10^{-5}$ for observable S-stars in our simulations.

In the right panel of Fig.~\ref{fig:rate_num}, we show the Hills mechanism rate against the predicted number of observed HVSs, main-sequence S-stars, and evolved S-stars. Given this star formation history, we expect, within our brightness limit, 7.2 and 0.05 main-sequence and evolved S-stars, respectively for every HVS in the \citetalias{Verberne_2024} survey. This means that the number of observed HVSs in \citetalias{Verberne_2024} is consistent with the twelve observed main-sequence S-stars within our brightness limit. If S5-HVS1 was ejected from a binary originating in the CWD, it is thus likely that most if not all main-sequence S-stars in the S-star cluster were deposited through the Hills mechanism. On the other hand, this model has trouble explaining the evolved S-stars, since for every evolved S-star it predicts around 140 main-sequence S-stars. This further demonstrates that if we assume the S-star cluster is exclusively formed through the Hills mechanism, the ratio of evolved to main-sequence S-stars is sensitive to the star formation history of the progenitor binary population. However, there is no obvious single progenitor population of binaries that can explain the number of HVSs, main-sequence S-stars, and evolved S-stars simultaneously. 

\subsection{Combined CWD and NSC progenitors}\label{sec:combined}
We argued in Section~\ref{sec:NSC}, that progenitor binaries exclusively originating from the NSC cannot explain the observed properties of HVSs and main-sequence/evolved S-stars. However, we would like to stress that theoretically, one would expect there to be a finite disruption rate $>0$ yr$^{-1}$ of binaries originating from the NSC. Furthermore, a young progenitor population is needed to produce sufficient main-sequence S-stars. A natural interpretation of the observed properties of the S-star cluster compared to our simulations is that the main-sequence S-stars were recently deposited on their current orbits by disruptions from a young population of stars, such as the CWD, while the evolved S-stars were deposited through disruptions of mainly old, low-mass binaries originating in, for example, the NSC. Here we investigate if this scenario is able to explain all the observables simultaneously.

\begin{figure}
    \centering
    \includegraphics[width=\linewidth]{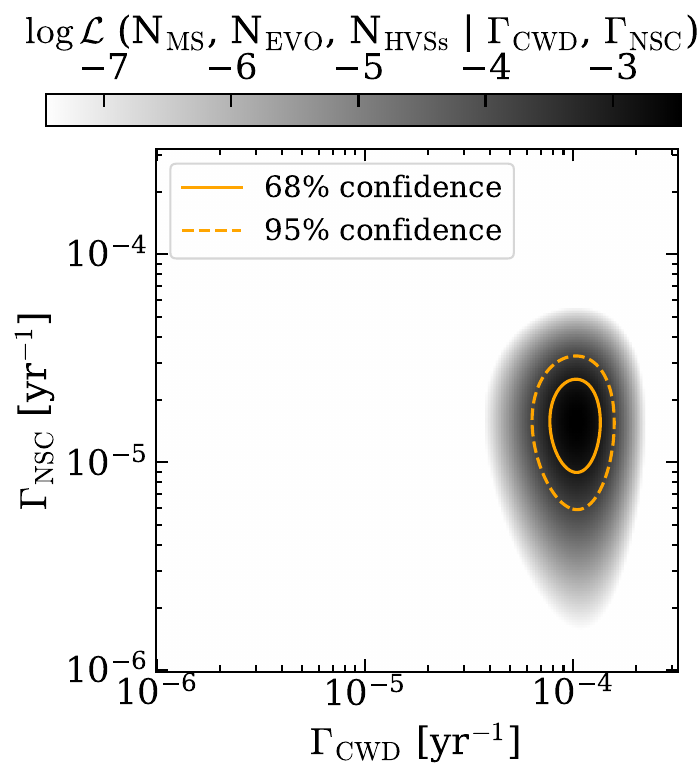}
    \caption{Log-likelihood on the Hills mechanism rate for our combined CWD and NSC model. In orange we show the 68th and 95th percentile confidence regions of our fit.}
    \label{fig:two-rates}
\end{figure}

If progenitor binaries are disrupted from both the NSC and CWD, we expect there to be two rates that make up the total Hills mechanism rate: a constant rate caused by the inflow of binaries from the NSC ($\Gamma_{\rm NSC}$) and a transient rate from the CWD ($\Gamma_{\rm CWD}$). Our model is a linear combination between these two rates. The three observables we fitted to are the number of HVSs, main-sequence S-stars, and evolved S-stars that are observed. We show the resulting log-likelihood of $\Gamma_{\rm CWD}$ and $\Gamma_{\rm NSC}$ in Fig.~\ref{fig:two-rates} in combination with the confidence regions of the 68th and 95th percentile.
We find that the rate of the Hills mechanism for binaries originating from the NSC is $1.4^{+0.9}_{-0.6}\times10^{-5}$ yr$^{-1}$ and that from the CWD is $1.0^{+0.3}_{-0.2}\times10^{-4}$ yr$^{-1}$. The current rate of Hills mechanism disruptions from the CWD would thus have to be about an order of magnitude higher than the constant background rate of binary disruptions from the NSC in order to fit the observables. Over the past $\sim10$ Myr, we therefore predict that $\sim90\%$ of ejected stars are young ($\lesssim10$ Myr). Our derived rates are comparable to old full and empty loss-cone rate estimates \citep{Hills_1988, Yu_2003}. In this scenario, S5-HVS1 is about 150 times more likely to have been ejected from the young progenitor population; this, combined with the age estimate for S5-HVS1 of less than $10^8$ yr \citep{Koposov_2020} and the alignment of the CWD plane with that of its progenitor binary, leads us to argue that S5-HVS1 came from the CWD and shares the same age.

\begin{figure*}
    \centering
    \includegraphics[width=0.85\linewidth]{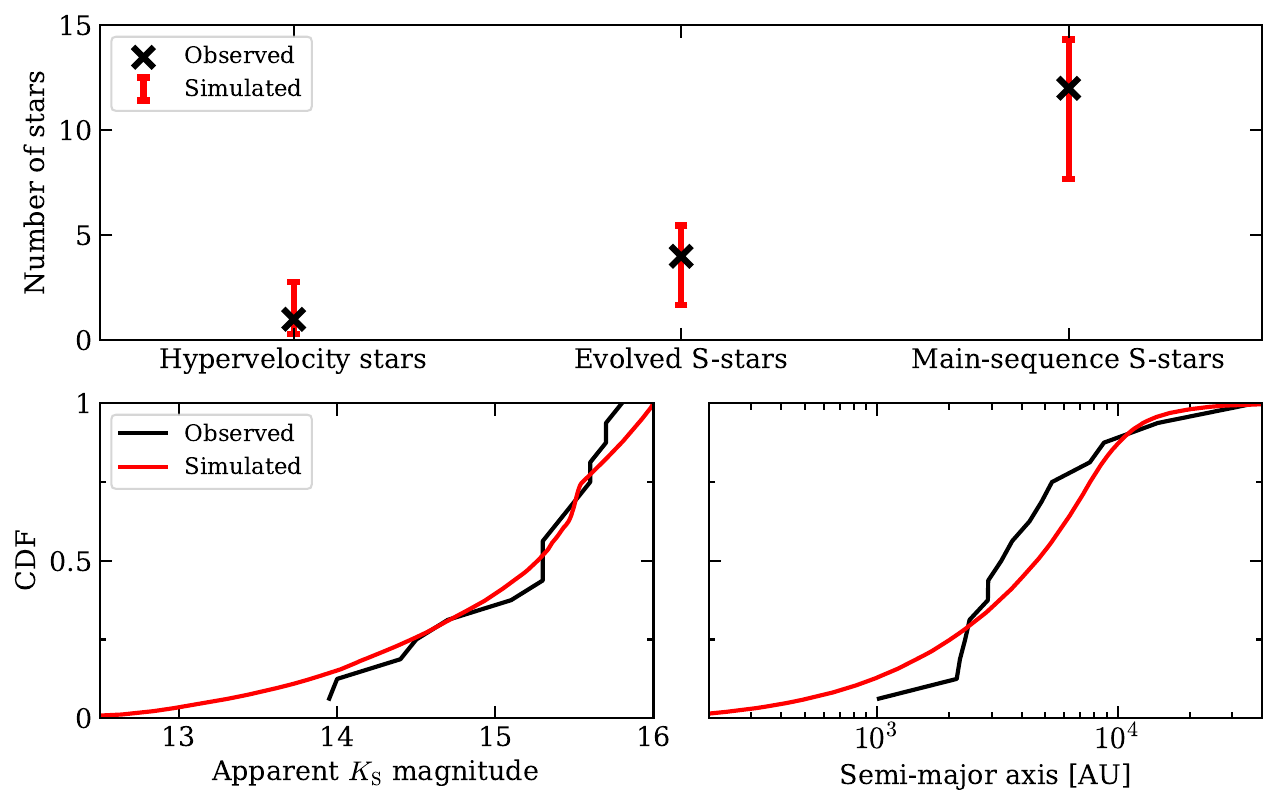}
    \caption{Comparison of our best fit model, including Hills mechanism disruptions from both the NSC and CWD, with observations. {\bf Top:} number of stars from our simulation compared to the observations for HVSs, evolved S-stars, and main-sequence S-stars. The error bars show the Poisson error on the predicted numbers from the model. {\bf Bottom left:} cumulative distribution function of the visual \ks magnitude of S-stars for our simulations compared to the observations. {\bf Bottom right:} cumulative distribution function of the semi-major axis of S-stars for our simulations compared to the observations.}
    \label{fig:combined_comparison}
\end{figure*}

Now that we have a combined CWD and NSC model, we evaluate if this model can reproduce all the observables of both the S-star cluster and HVSs in Fig.~\ref{fig:combined_comparison}.
Firstly, we recount that neither progenitor population, when considered individually, was able to simultaneously predict the observed numbers of HVSs, main-sequence S-stars, and evolved S-stars. However, as shown in the top panel of Fig.~\ref{fig:combined_comparison}, the combined model successfully accounts for all three of these quantities at once. Moreover, both the \ks-band and semi-major axis distributions are consistent between our model and the observations, with KS-test p-values of 0.88 and 0.55 respectively.

\subsubsection*{Stellar tidal disruption events}
We find that for both our progenitor populations about 20\% of Hills mechanism disruptions result in a stellar tidal disruption event (TDE). The resulting TDE rate of the order $10^{-5}$ to $10^{-6}$ yr$^{-1}$ is consistent with extragalactic TDE rates \citep{Stone_2016}. However, this estimate is strongly influenced by our assumption of a full loss-cone. In the empty loss-cone regime, the Hills mechanism contribution to the TDE rate would be about two orders of magnitude lower. Our estimated Hills mechanism contribution to the TDE rate of $10^{-5}$ yr$^{-1}$ should thus be seen as an upper limit.

\section{Discussion}
\label{sec:discussion}
In this work we have investigated whether the Hills mechanism can explain both the observed properties of the S-star cluster and HVSs, and we found a plausible formation scenario that is consistent with data (see Section~\ref{sec:combined} and Fig.~\ref{fig:combined_comparison}). In this Section we critically review our model assumptions in the first five subsections. We investigate further the comparison between the semi-major axis distribution of our simulated population and the observed one in Section~\ref{sec:asm}. We discuss future prospects in Section~\ref{sec:prospects}.

\subsection{Star formation history}\label{sec:star_formation_history}
\begin{figure}
    \centering
    \includegraphics[width=\linewidth]{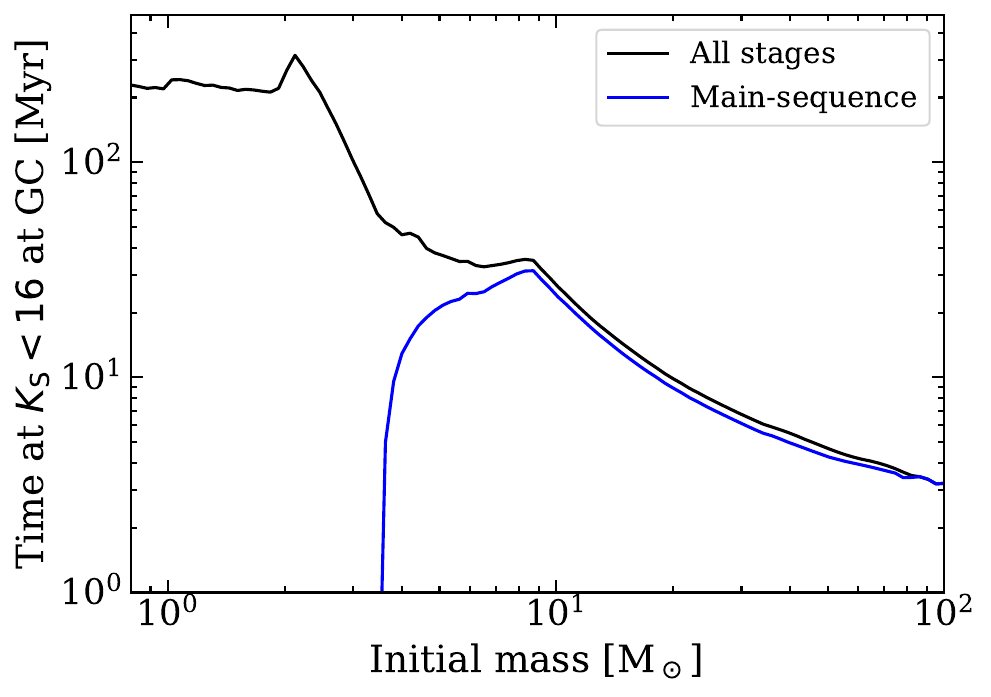}
    \caption{Initial stellar mass against the time spent with a magnitude of less than 16 in the \ks band, when the star is located at the GC. We utilise the stellar evolution code \texttt{MIST}.}
    \label{fig:time_bright}
\end{figure}

In this work, we have investigated two progenitor binary populations. The most important parameter that sets the ratio of main-sequence to observed S-stars is the star formation history. To demonstrate this, we calculated the amount of time a star in the GC can be above our brightness limit as a function of its initial mass, and we show this in Fig.~\ref{fig:time_bright}.
First of all, we notice that only main-sequence stars more massive than about 4 M$_\odot$ can be observed out to the GC, which have a main-sequence lifetime of $\lesssim170$ Myr. Secondly, we notice that evolved, low-mass stars can be brighter than our threshold for about an order of magnitude longer than massive main-sequence stars. For an old population of stars, such as the NSC, the result will thus be that most of the stars above the brightness limit will be evolved, low-mass stars. For a young population, such as the CWD, most of the stars above the brightness limit will instead be massive main-sequence stars. The ratio of evolved to main-sequence stars therefore traces the star formation history.

Our results show that a young progenitor population is needed to produce sufficient main-sequence S-stars to match observations. Our conclusion here is in line with \citet{Generozov_2024}, who, through a different argument, recently found the need for a young population of stars to explain the main-sequence S-stars. However, we argue that binary disruptions exclusively from a young population such as that of the CWD cannot explain the presence of the evolved S-stars. 

Our conclusions most importantly relate to the star formation history of the progenitor populations. Alternative populations of possible old progenitor binaries, such as the nuclear stellar disc \citep{Schodel_2023}, result in similar conclusions. This is particularly relevant, since massive perturbers might increase the disruption rate of binaries originating at these radii by several orders of magnitude \citep{Perets_2007}. In the case of the young progenitor binary population, we consider the CWD to be the most likely origin, given its close proximity to Sgr A* and its young age. 

As a more agnostic star formation history, we also investigated a constant star formation rate over the past 13 Gyr and a canonical IMF slope of $-2.3$ for our progenitor population. Based on these simulations, we expect there to be 0.11 and $1.5\times10^2$ main-sequence and evolved S-stars respectively within the brightness limit for every HVS in the \citetalias{Verberne_2024} survey. This is qualitatively similar to the results for the NSC progenitor origin and also incompatible with observations. \\

In the context of star formation history, we would also like to point out that the commonly used single power-law slope for the (present-day) mass function implicitly assumes a constant star formation rate with an IMF that is not explicitly defined. Since the stellar lifetime roughly scales with $M^{-2.5}$ \citep[e.g.][]{ryan2010stellar}, the IMF can be approximated by subtracting about $-2.5$ from the power-law slope of the current mass function if star formation was constant over the last several billion years. For a population of stars with a current mass function slope of $-2.3$, the IMF would thus be about $\frac{dN}{dM}\propto M^{0.2}$, unless star formation was not constant. Such a highly top-heavy IMF is not realistic even for the top-heavy CWD \citep[e.g.][]{Bartko_2010, Lu_2013, Gallego-Cano_2024}.

\subsection{Stellar collisions}\label{sec:collisions}
So far, we have ignored the effects of stellar collisions. In general, in a dense cluster of stars and black holes with a high velocity dispersion, like that occupied by the S-stars, stellar collisions might become important. There is currently no consensus as to the importance of collisions for shaping the S-star cluster. Collisions may shape the density profiles of NSCs on a scale similar to that of the S-star cluster, depleting stars at small radii and potentially even leading to mergers \citep{Rose_2023, Rose_2024, Ashkenazy_2024}. \citet{Generozov_2024}, however, find that collisions only have a minor impact on their results.

It should also be noted that there is no consensus as to the net effect of collisions. Depending on the mass ratio, evolutionary stage of each component involved, impact parameter, and impact velocity the result of a collision will vary drastically \citep[e.g.][]{Benz_1987, Trac_2007, Mastrobuono-Battisti_2021, Rose_2023, Rose_2024}. 

\begin{figure}
    \centering
    \includegraphics[width=\linewidth]{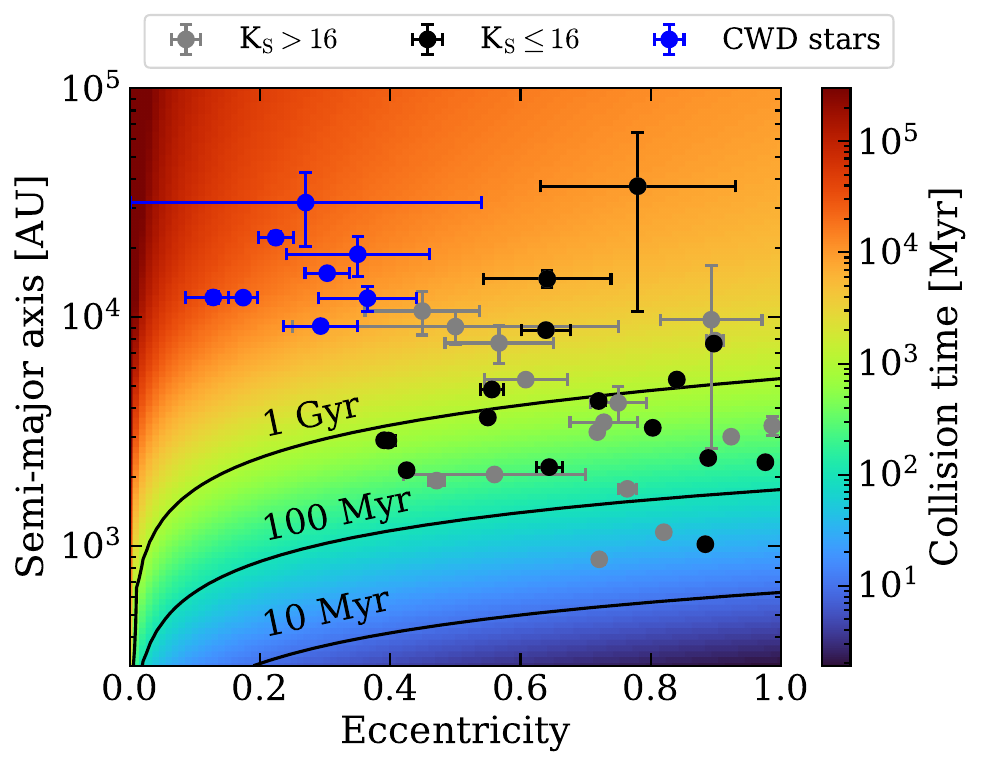}
    \caption{Collisional timescale of a 10 M$_{\odot}$ star with a 5.6 R$_{\odot}$ radius as a function of stellar eccentricity and semi-major axis. The stars with orbital solutions (see Fig.~\ref{fig:S-stars}) are overplotted.}
    \label{fig:collisions}
\end{figure}

In Fig.~\ref{fig:collisions}, we show the collisional timescale as a function of eccentricity and semi-major axis for a \citet{Bahcall_1976} cusp following \citet{Sari_2019}.
We can see that stellar collisions might be important for a significant part of the parameter space. Particularly for low-mass stars that could otherwise reside in the S-star cluster for several billion years during their main-sequence, collisions might play an important role. The massive main-sequence S-stars will not be significantly affected, since their main-sequence lifetimes tend to be shorter than the collisional timescale shown here. 

Collisions are thus relevant for our NSC progenitor population model. To ensure our conclusions are robust, we also calculated our NSC progenitor results for ejections only over the past 100 Myr. The effect of only considering recent ejections is that the predicted number of evolved S-stars decreases. However, even if we only consider these recent binary disruptions, we would still expect about 24 evolved S-stars for every main-sequence S-star. This means that even when only considering recent disruptions, an old progenitor binary population is inconsistent with observations. Because possible depletion will remove a fraction of the simulated S-stars, we expect that the Hills mechanism rate constraint for the NSC in our combined model is a lower limit. The true rate would need to be higher to account for the depletion rate over time.

\subsection{Metallicity}\label{sec:metallicity}
The metallicity of the S-stars is, as of yet, unknown \citep{Habibi_2017}. For this reason, it is unclear what metallicity to assume for the progenitor binary population.

In the case of the NSC, most studies find a super-solar mean metallicity of \feh$\sim 0.2$ to 0.3 \citep[e.g.][]{Schultheis_2019, Schodel_2020} with a minor population of a few percent being metal poor \citep[e.g.][]{Do_2015, Feldmeier-Krause_2017, Gallego-Cano_2024}. The metallicity of the CWD, like that of the S-stars, is unknown, though in situ star formation in the GC is generally expected to be (super) solar \citep[e.g.][]{Do_2015, Feldmeier-Krause_2017, Schodel_2020}. 

Fortunately, the effect of the assumed metallicity in this study is minimal, since we only consider if stars are on the main-sequence or evolved, in combination with a brightness limit. If we assume a metallicity of \feh$=0.3$, we find that the number of main-sequence S-stars to evolved stars changes from 7.0 to 7.5 and the number of evolved S-stars from 0.05 to 0.04 for the CWD binary origin. Also for the HVS observations, metallicity is unimportant, because of the large coverage of colour-magnitude space \citepalias{Verberne_2024}.

\subsection{Energy diffusion}\label{sec:E_diff}
\begin{figure}
    \centering
    \includegraphics[width=\linewidth]{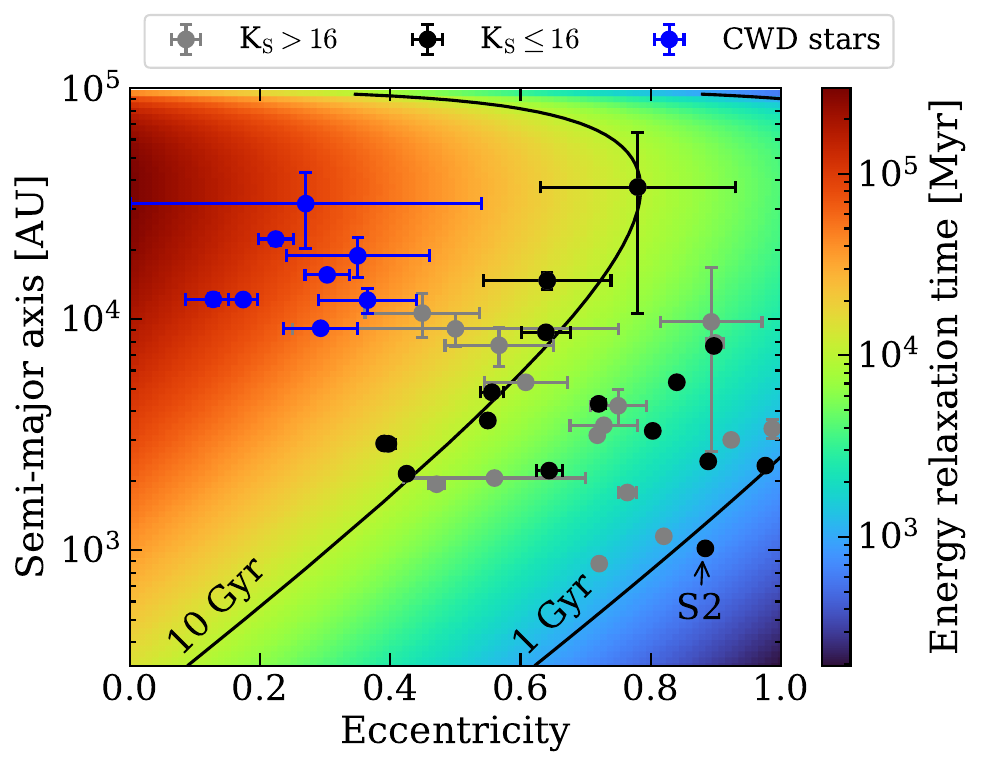}
    \caption{Two-body energy relaxation timescale as a function of eccentricity and semi-major axis assuming a \citet{Bahcall_1976} cusp. The stars with orbital solutions are overplotted (see Fig.~\ref{fig:S-stars}). The two solid lines mark the 1 and 10 Gyr relaxation times.}
    \label{fig:two-body}
\end{figure}

In this paper, we assumed that the energy of any captured stars does not evolve after the tidal interaction. In reality, the energy diffuses on the two-body relaxation timescale \citep[e.g.][]{Bar-Or_2016, Sari_2019}. If we use the formalism from \citet{Sari_2019}, while assuming a \citet{Bahcall_1976} cusp of stars, we get the two-body relaxation timescale in energy shown in Fig.~\ref{fig:two-body}.
We can see that only S2 has a shorter energy relaxation time than 1 Gyr, which is still much longer than its main-sequence lifetime \citep{Habibi_2017}. We noted in Section~\ref{sec:star_formation_history}, that only main-sequence stars with masses above $\sim4$ M$_\odot$ are expected to pass our brightness limit, for which the main-sequence lifetime is about 170 Myr. This is much shorter than the typical two-body relaxation timescale for the energy shown in Fig.~\ref{fig:two-body}. We conclude that for the assumed density profile, our assumption that the captured stars' energy is constant is safe for main-sequence stars. If, instead, a star was deposited $>1$ Gyr ago, the energy might have changed significantly. This is the case for very old disruptions from the NSC progenitor population discussed in Section~\ref{sec:NSC}.

\subsection{Progenitor binary eccentricity}\label{sec:eccentricity}
The internal eccentricity of progenitor binaries influences the energy that is exchanged in the Hills mechanism through the prefactor that we plot in Fig.~\ref{fig:alpha}. If we assume a thermal eccentricity distribution for our progenitor binaries originating in the CWD, the number of main-sequence and evolved S-stars per observed HVS change from 7.2 and 0.05 to 7.1 and 0.06 respectively. Moreover, if we compare the cumulative distribution functions of the simulated semi-major axes of the S-stars, we find a maximum difference of about 0.02, which is well below statistical uncertainties given the number of observed S-stars. Assuming a thermal eccentricity distribution or a circular one makes thus little difference for our results. This is explained by considering that disruptions that result in observable S-stars tend to be of high energy, which means high values for the prefactor. For high prefactors, we can see from Fig.~\ref{fig:alpha} that the thermal eccentricity distribution is very similar to the circular one.

\subsection{The semi-major axis distribution}\label{sec:asm}
So far, we have not discussed the semi-major axis distribution of observed and modelled S-stars in-depth. The reason is that it is the most difficult observable to interpret. One needs to model the impact of collisions in combination with energy and angular momentum diffusion and observational biases, before attempting to compare a simulated semi-major axis distribution with an observed one.

We showed in Fig.~\ref{fig:combined_comparison} how our model compares to the observed semi-major axis distribution. We found the two to be consistent, but the relatively limited number of S-stars for which we can assume the observations to be complete limits our sensitivity. The lack of predicted S-stars at semi-major axes greater than about $2\times10^4$ AU is not surprising, since this is caused by the observational selection function \citep{Burkert_2024}. However, our model is able to naturally explain the lack of S-stars at semi-major axes below about $10^3$ AU, without the need for stellar collisions. The limiting factors for producing stars at such radii are the energy that can be exchanged in the Hills mechanism and the high initial eccentricity that can cause TDEs to occur for small semi-major axes.

\begin{figure}
    \centering
    \includegraphics[width=\linewidth]{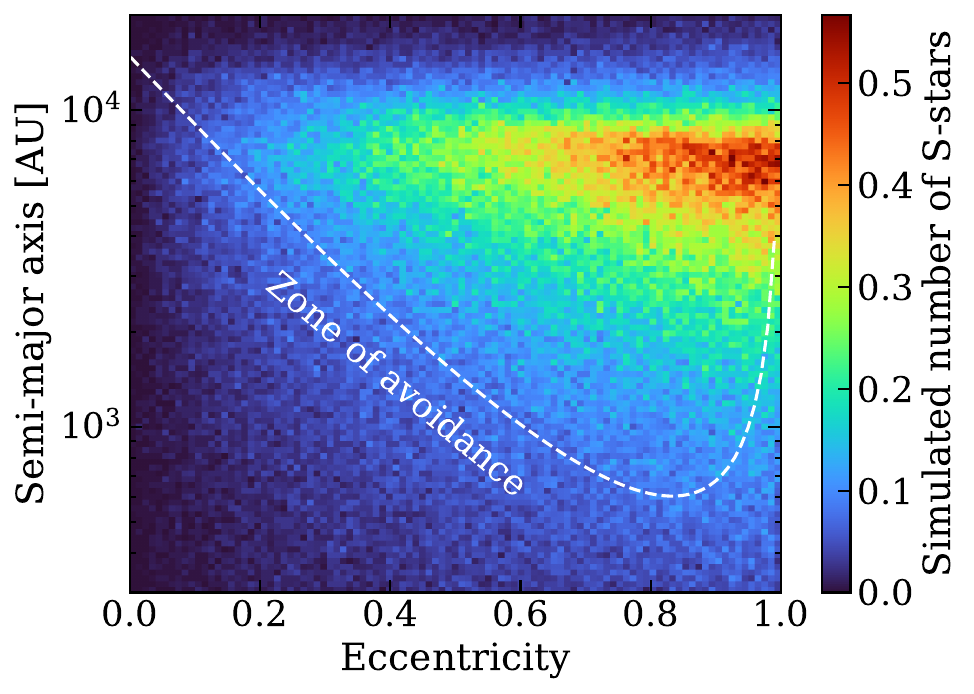}
    \caption{Eccentricity against semi-major axis for our simulation of the CWD origin binaries. The colour bar shows the number of S-stars predicted for an ejection rate of $10^{-2}$ yr$^{-1}$. In white we show the boundary of the zone of avoidance identified in \citet{Burkert_2024}. The drop in the number of simulated S-stars at large semi-major axes is due to the selection function from \citet{Burkert_2024}.}
    \label{fig:zone_of_avoidance}
\end{figure}

In the literature, a 'zone of avoidance' was identified in pericentre distance against eccentricity where no S-stars have been found, even though observations should have identified them if they exist. \citet{Burkert_2024} defines the region by $\log(r_{\rm p})<1.57 + 2.6\times(1-e)$, where $r_{\rm p}$ and $e$ are the pericentre distance and eccentricity respectively. In Fig.~\ref{fig:zone_of_avoidance}, we show the zone of avoidance in semi-major axis and eccentricity space (which are independent quantities), in combination with the predicted S-star cluster distribution of our CWD origin simulations.
The figure demonstrates that our simulations can naturally explain the lack of observed stars in the zone of avoidance. The fundamental reasons being the limit on the energy that can be exchanged in the Hills mechanism (equation~\ref{eq:delta_E}), captured stars ending up as TDEs at small semi-major axis  (since their initial eccentricity is close to 1), and subsequent relaxation of the eccentricity distribution. Our results therefore agree with the conclusion of \citet{Generozov_2024}, that the Hills mechanism can naturally explain the lack of orbital solutions in the zone of avoidance.

\subsection{Prospects}\label{sec:prospects}
Definitively identifying the formation origin of the S-star cluster is very challenging. Key observables of the S-stars that could assist in this are in particular the individual ages, metallicities, and masses of the S-stars. These quantities could help distinguish if the S-stars were formed from multiple populations of stars, or a single star formation episode, which is especially useful for evaluating scenarios. Of particular interest for the Hills mechanism origin hypothesis is the age and metallicity distribution of the evolved S-stars. If the evolved S-stars are old, low-mass stars it would suggest that the main-sequence stars have a different origin. Either the Hills mechanism efficiently disrupts binaries from young structures such as the CWD or alternative scenarios are needed to explain their presence near Sgr A*, such as those explored by \citet{Levin_2007}, \citet{Rantala_2024}, and \citet{Akiba_2024}.

A deeper completeness limit of S-stars observed near Sgr A* would also help eliminate formation scenarios, since it would allow for more detailed comparisons using the \ks-band luminosity function and semi-major axis distributions in particular. This deeper completeness limit will be allowed by new instruments such as GRAVITY(+) \citep{Eisenhauer_2011, Gravity+_2022}, and an increased temporal baseline for the observations.

Specifically for relating the Hills mechanism to the S-star cluster, a larger sample of confirmed HVSs originating in the GC is vital. This would allow for much lower uncertainties in the Hills mechanism rate, even for a single new HVS discovery. Additionally, measurements of the metallicities, ages, and masses of HVSs would provide strong constraints on their progenitor population. Large spectroscopic surveys such as \gaia DR4, {\it DESI} \citep{Cooper_2023}, {\it WEAVE} \citep{Dalton_2014}, and {\it 4MOST} \citep{deJong_2019} will provide unprecedented numbers and depth of observations and are expected to discover new HVSs \citepalias{Verberne_2024}.

\section{Conclusions}\label{sec:conclusions}
In this work, we investigated if binary disruptions through the Hills mechanism can explain both the observed population of S-stars and state-of-the-art HVS observations. We forward modelled the S-stars and HVSs and compared both populations to the latest observational results, taking observational selection effects into account. We find that, in particular, the ratio of main-sequence to evolved S-stars is highly informative on the star formation history of the progenitor binary population and investigated different progenitor binary populations. 

We show that an old progenitor binary population, where the majority of the stellar content formed billions of years ago such as the NSC, cannot explain the ratio of main-sequence to evolved S-stars, nor the relative number of HVSs. Furthermore, our results show that a young progenitor binary population, such as the CWD, can simultaneously explain the observed population of main-sequence S-stars and HVS observations, but cannot explain the evolved S-stars that are observed. 

We argue that the main-sequence and evolved S-star populations, if formed through the Hills mechanism, have different progenitor populations. Only the model in which progenitor binaries come from both the CWD and NSC can successfully the numbers of HVSs, main-sequence S-stars, and evolved S-stars. Furthermore, we show that the \ks-band and semi-major axis distributions predicted by this model are consistent with observations. The current Hills mechanism rate from the young binary population would need to be about an order of magnitude higher than that of the old population at roughly $10^{-4}$ and $10^{-5}$ yr$^{-1}$ respectively. About 90\% of ejected stars over the past $\lesssim10$ Myr should thus originate from the CWD. In general, we expect most young HVSs to be ejected in bursts coinciding with star formation near Sgr A*.

We additionally argue that, provided S5-HVS1 was formed through the Hills mechanism, most (if not all) S-stars were formed through Hills mechanism disruptions of binaries.

Upcoming large spectroscopic surveys are expected to improve our understanding of the rate of the Hills mechanism, and thereby provide strong constraints on the Hills mechanism's contribution to the formation of the S-star cluster. In addition, continued observations of the GC will allow more precise measurements of the properties of the S-stars, facilitating increasingly constraining tests of different formation scenarios.

\begin{acknowledgements}
The authors thank Stephan Gillessen, Antonia Drescher, Diogo Ribeiro, Matteo Bordoni, Bianca Sersante, and Yuri Levin for helpful discussions.
EMR acknowledges support from European Research Council (ERC) grant number: 101002511/project acronym: VEGA\_P. SK acknowledges support from the Science \& Technology Facilities Council (STFC) grant ST/Y001001/1. 
For the purpose of open access, the author has applied a Creative
Commons Attribution (CC BY) licence to any Author Accepted Manuscript version
arising from this submission.\\

Software: \texttt{NumPy} \citep{Harris_2020}, \texttt{SciPy} \citep{2020SciPy-NMeth}, \texttt{Matplotlib} \citep{Hunter_2007}, \texttt{Astropy} \citep{astropy:2013, astropy:2018, astropy:2022}, \texttt{isochrones} \citep{Isochrones_2015}, \texttt{Speedystar} \citep{Contigiani_2019, Evans_2022_I}.
\end{acknowledgements}

\bibliographystyle{aa}
\bibliography{mybib}

\end{document}